# *BVRI* Photometric Observations, Light Curve Solutions and Orbital Period Analysis of BF Pav

Atila Poro[1,2], Fahri Alicavus[3,4], Eduardo Fernández-Lajús[5,6], Fatemeh Davoudi[1,2], PegahSadat MirshafieKhozani[1], Mark G. Blackford[7], Edwin Budding[8], Behjat Zarei Jalalabadi[1], Jabar Rahimi[1], Farzaneh Ahangarani Farahani[1]

[1]The International Occultation Timing Association Middle East section, Iran, info@iota-me.com
[2]Astronomy Department of the Raderon Lab., Burnaby, BC, Canada
[3]Çanakkale Onsekiz Mart University, Faculty of Arts and Sciences, Department of Physics, 17020, Çanakkale, Turkey
[4]Çanakkale Onsekiz Mart University, Astrophysics Research Center and Ulupınar Observatory, 17020, Çanakkale, Turkey
[5]Instituto de Astrofísica de La Plata (CCT La Plata-CONICET-UNLP), La Plata, Argentina
[6]Facultad de Ciencias Astronómicas y Geofísicas, Universidad Nacional de La Plata, Paseo del Bosque, B1900FWA, La Plata, Argentina
[7]Variable Stars South (VSS), Congarinni Observatory, Congarinni, NSW, 2447, Australia
[8]Carter Observatory, 40 Salamanca Rd, Kelburn, Wellington 6012, New Zealand

**Abstract**
A new ephemeris, period change analysis and light curve modeling of the W UMa-type eclipsing binary BF Pav are presented in this study. Light curves of the system taken in BVRI filters from two observatories in Australia and Argentina were modeled using the Wilson-Devinney code. The results of this analysis demonstrate that BF Pav is a contact binary system with a photometric mass ratio $q = 1.460 \pm 0.014$, a fillout factor $f = 12.5\%$, an inclination of $87.97 \pm 0.45 \ deg$ and a cold spot on the secondary component. By using the distance modulus formula, the distance of BF Pav was calculated to be $d = 268 \pm 18 \ pc$ which is in good agreement with the Gaia EDR3 distance. We obtain an orbital period increase at a rate of 0.142 s/century due to a quadratic trend in O-C diagram. Also, an alternative sudden period jump probably has occurred which could be interpreted as a rapid mass transfer from the lower mass star to its companion about $\Delta M = 2.45 \times 10^{-6} \ M_\odot$. Furthermore, there is an oscillatory behavior with a period of $18.3 \pm 0.3 \ yr$. Since BF Pav does not seem to have significant magnetic activity, this behavior could be interpreted as the light-time effect caused by an undetected third body in this system. In this case, the probability for the third body to be a low mass star with $M >= 0.075 \ M_\odot$ or a brown dwarf is 5.4% and 94.6% respectively. If we assume $i' = 90°$, $a_3 = 8.04 \pm 0.33 \ AU$. The mass of the secondary component was also determined using two different methods which result close to each other.

## 1. INTRODUCTION

W UMa-type binary systems have short orbital periods less than a day and they show continuous light variations (Dryomova & Svechnikov 2006). These systems are abundant in binary stars (Okamoto & Sato 1970). They include two stars usually surrounded by a common envelope resulting from a mass overflowing from the Roche lobe of one binary component (Smith 1984). Despite many studies that have been done on the basis of the Common Connective Envelope (CCE) in recent years (Qian 2003), many details still undetermined about the evolutionary state of the contact binaries due to extreme spectral line broadening for achieving spectra analysis (Yang et al. 2015). For justifying the contact phase in contact binaries, Stępień (2006) suggested the angular momentum loss through the magnetic wind, whereas Qian et al. (2018) suggested the transmission of a large extent of angular momentum to a third body. So it is efficient to investigate the formation and structure of W Uma-type contact binaries for studying physical processes in these systems with more details. We can find valuable details about mass transfer, mass loss, and also the evolutionary state of close binaries by perusing orbital variations of them.

The BF Pav binary system, which is located in the constellation of Pavonis in the Southern Hemisphere Sky, is a variable star of W UMa-type with an approximate period of 0.30231864 days and G8 Spectral type (Gonzalez et al. 1996). Its apparent magnitude in the *V*-band is 12.17 (APASS9). The variability of BF Pav was discovered by Shapley in 1939 and the first photoelectric light curve was obtained by Hoffman (1981). Although these observations did not cover the complete orbital period, the observer derived a period of 0.3056 days (Gonzalez et al. 1996). Between 1987 and 1993, BF Pav was observed photoelectrically in the *UBV* filters in the observational program of Southern Short-Period Eclipsing Binaries to determine the times of minima, photometric and absolute parameters. The photometric solution resulted with a mass ratio of $q = 1.4 \pm 0.2$, a fillout factor equals to 10%, and efficient thermal contact between the components, $\Delta T \ 100K$ (Gonzalez et al. 1996). Dryomova and Svechnikov (2006) found the rate of period change of BF Pav to be $\dot{p} = 1.62 \times 10^{-7} \frac{days}{year}$ in their study by checking the variation in the orbital period of W Uma-type contact systems. Zhang et al. (2015) noted that BF Pav has a similar period increase to GK Aqr.

In this paper, we present a new ephemeris based on our observations as well as new period change analysis and light curve solutions to investigate the evolutionary state of BF Pav in more detail.

## 2. OBSERVATION AND DATA REDUCTION

The observation of the BF Pav was carried out in September 2017, April 2018, August and June 2019, and July 2020, and a total of 3517 images were taken during eight nights. 2054 images were taken with a 14-inch Ritchey Chretien telescope and SBIG STT3200-ME CCD equipped with Astrodon Johnson-Cousins *BVRI* Filters at the Congarinni Observatory which is located in Australia with geographical coordinates 152° 52´ East and 30° 44´ South and 20 meters above the mean sea level. Each frame was recorded at $2 \times 2$ binning with 50 seconds exposure time in each filter and CCD temperature set at -15°C. Another 1463 images were taken with the 2.15 m "Jorge Sahade" telescope at the Complejo Astronomico El Leoncito (CASLEO) Observatory (69° 18´ W, 31° 48´ S, 2552 m above the sea level), Argentina. A VersArray 2048B, Roper Scientific cryogenic CCD, and a *V*-band filter were used. Each frame was recorded at $5 \times 5$ binning with 15 s exposure time.

GSC 8770-1511 was chosen as a check star and 8 stars were chosen as comparison stars with appropriate apparent magnitude in comparison to BF Pav. The general characteristics of BF Pav with the comparisons and the check star are shown in Table 1.

Table 1 Characteristic of the Variable star, the Check star, and the Comparison stars (from: SIMBAD[1] and APASS9[2]).

| Star type | Star name | RA. (J2000) | DEC. (J2000) | Magnitude (*V*) |
|---|---|---|---|---|
| Variable | BF Pav | 18 45 39.32 | -59 38 25.87 | 12.17 |
| Comparison$_1$ | GSC 8770-1107 | 18 45 30.01 | -59 32 34.9 | 12.23 |
| Comparison$_2$ | GSC 8770-1582 | 18 45 50.08 | -59 36 59.1 | 13.43 |
| Comparison$_3$ | GSC 8770-1663 | 18 45 49.66 | -59 37 48.9 | 13.43 |
| Comparison$_4$ | GSC 8770-0085 | 18 45 33.39 | -59 39 50.5 | 13.52 |
| Comparison$_5$ | GSC 8770-103 | 18 46 0.36 | -59 36 5.8 | 12.18 |
| Comparison$_6$ | GSC 8770-1383 | 18 45 42.38 | -59 32 5.3 | 13.39 |
| Comparison$_7$ | GSC 8770-1325 | 18 45 18.25 | -59 44 27.2 | 13.23 |
| Comparison$_8$ | GSC 8770-1333 | 18 45 56.89 | -59 40 26.6 | 13.54 |
| Check | GSC 8770-1511 | 18 45 32.73 | -59 36 25.0 | 13.48 |

Standard procedures for CCD image processing (aligned pictures, bias and dark removal, flat-fielding to correct for vignetting, and pixel-to-pixel variations) were applied. We did all image processing and plotting raw images with MaxIm DL software (George 2000). Then more modifications were made with AstroImageJ (AIJ) software (Collins et al. 2017). AIJ is a powerful tool for astronomical image analysis and precise photometry (Davoudi et al. 2020).

We determined 11 primary and 8 secondary minimum times from the observed light curves in *BVRI* filters. These minima were calculated by using the Kwee and van Woerden (1956) method.

---

[1] http://simbad.u-strasbg.fr/simbad/
[2] http://vizier.u-strasbg.fr/viz-bin/

## 3. ORBITAL PERIOD VARIATIONS

Using 58 mid-eclipse times including 30 primary and 28 secondary eclipses from the previous study and our observations, we analyzed the orbital period variation of this system. We averaged all the times of minima from literature and our observation that were in the same filter at the same time. All times of minimum are expressed in Barycentric Julian Date in Barycentric Dynamical Time (BJD$_{TDB}$) and listed in Table 2. It includes errors, epochs, O-C values, and the references of mid-eclipse times in the last column. The linear ephemeris of Gonzalez et al. (1996) was used for computing epochs and the O-C values,

$$Min.I(BJD_{TDB}) = 2448056.9014(\pm0.0002) + 0.30231864(\pm0.00000007) \times E. \quad (1)$$

Table 2 Available times of minima for BF Pav.

| Min. (BJD$_{TDB}$) | Error | Epoch | O-C | References |
|---|---|---|---|---|
| 2444438.7617 | | -11968 | 0.0098 | (Hoffmann 1981) |
| 2445886.4094 | 0.0001 | -7179.5 | 0.0047 | (Jones 1988) |
| 2445886.5621 | 0.0004 | -7179 | 0.0062 | (Jones 1988) |
| 2446936.8129 | | -3705 | 0.0021 | (Gonzalez et al. 1996) |
| 2447259.8388 | | -2636.5 | 0.0005 | (Gonzalez et al. 1996) |
| 2447368.6721 | | -2276.5 | -0.0009 | (Gonzalez et al. 1996) |
| 2448056.7488 | | -0.5 | -0.0014 | (Gonzalez et al. 1996) |
| 2448056.9009 | | 0 | -0.0005 | (Gonzalez et al. 1996) |
| 2448056.9014 | 0.0002 | 0 | 0.0000 | (Gonzalez et al. 1996) |
| 2448057.8076 | | 3 | -0.0008 | (Gonzalez et al. 1996) |
| 2448058.8660 | | 6.5 | -0.0005 | (Gonzalez et al. 1996) |
| 2449182.7360 | | 3724 | 0.0000 | (Gonzalez et al. 1996) |
| 2449184.7009 | | 3730.5 | -0.0002 | (Gonzalez et al. 1996) |
| 2449217.5031 | | 3839 | 0.0004 | (Gonzalez et al. 1996) |
| 2449217.6548 | | 3839.5 | 0.0010 | (Gonzalez et al. 1996) |
| 2449218.5615 | | 3842.5 | 0.0007 | (Gonzalez et al. 1996) |
| 2449219.6190 | | 3846 | 0.0001 | (Gonzalez et al. 1996) |
| 2452404.5462 | 0.0001 | 14381 | 0.0004 | Zakrzewski (ASAS-3) |
| 2452404.6964 | 0.0002 | 14381.5 | -0.0005 | Zakrzewski (ASAS-3) |
| 2453479.4393 | 0.0005 | 17936.5 | -0.0004 | Zakrzewski (ASAS-3) |
| 2453558.4955 | 0.0004 | 18198 | -0.0005 | Zakrzewski (ASAS-3) |
| 2454606.9368 | 0.0003 | 21666 | -0.0003 | Zakrzewski (ASAS-3) |
| 2454614.9485 | 0.0001 | 21692.5 | 0.0000 | Zakrzewski (ASAS-3) |
| 2454778.6540 | 0.0001 | 22234 | 0.0000 | Zakrzewski (CATALINA) |
| 2454921.4991 | 0.0002 | 22706.5 | -0.0005 | Zakrzewski (CATALINA) |
| 2457172.7180 | 0.0004 | 30153 | 0.0026 | (Juryšek et al. 2017) |
| 2458015.5829 | 0.0001 | 32941 | 0.0032 | This study |
| 2458227.8106 | 0.0001 | 33643 | 0.0032 | This study |
| 2458639.7184 | 0.0001 | 35005.5 | 0.0018 | This study |
| 2458639.8699 | 0.0004 | 35006 | 0.0022 | This study |
| 2458701.9957 | 0.0010 | 35211.5 | 0.0015 | This study |
| 2458702.1472 | 0.0010 | 35212 | 0.0018 | This study |
| 2458710.9145 | 0.0010 | 35241 | 0.0019 | This study |
| 2458711.0654 | 0.0010 | 35241.5 | 0.0016 | This study |
| 2458713.9379 | 0.0001 | 35251 | 0.0021 | This study |
| 2458714.0884 | 0.0001 | 35251.5 | 0.0015 | This study |
| 2458716.9609 | 0.0001 | 35261 | 0.0019 | This study |
| 2458717.1116 | 0.0001 | 35261.5 | 0.0015 | This study |
| 2459060.9991 | 0.0002 | 36399 | 0.0015 | This study |

Note: Eight unpublished minimum times were provided by B. Zakrzewski and were determined from ASAS-3 and CATALINA Sky Survey data, according to the Timing Database at Krakow (Kreiner 2004).

To proceed with the analysis of the behavior of the period, we first averaged all the minimum time of Table 2 that correspond to the same event, in such a way that they do not have an overestimated weight in the adjustments. In the cases where an estimate of error was not reported in the original references, we assumed the error as the order of its last significant digit. Then, a first quadratic fitting using the least-squares method was done, using the errors to weight properly each data point considering ($w = \frac{1}{err^2}$). The following ephemeris formula was obtained,

$$Min. I\ (BJD_{TDB}) = (2448056.90186 \pm 0.00027) + (0.30231846 \pm 5.0 \times 10^{-8}) \times E + ((6.8 \pm 1.5) \times 10^{-12}) \times E^2 [days] \quad (2)$$

where $E$ is the cycle number after the reference cycle. The rate of change of the binary period is given by the quadratic coefficient "$Q$" of Equation 2, as follows,

$$\frac{dp}{dE} = 2Q \quad (3)$$

$\frac{dp}{dE} = (3.6 \pm 3.0) \times 10^{-12}$ day/cycle which represents a continuous period increase at a rate of $(16.4 \pm 3.6) \times 10^{-8}$ d/yr or 0.142 s/century. Figure 1 shows the O-C diagram calculated using Equation 2 and the residuals of the fitting. The solid black line represents the quadratic least-squares fit to the O-C values.

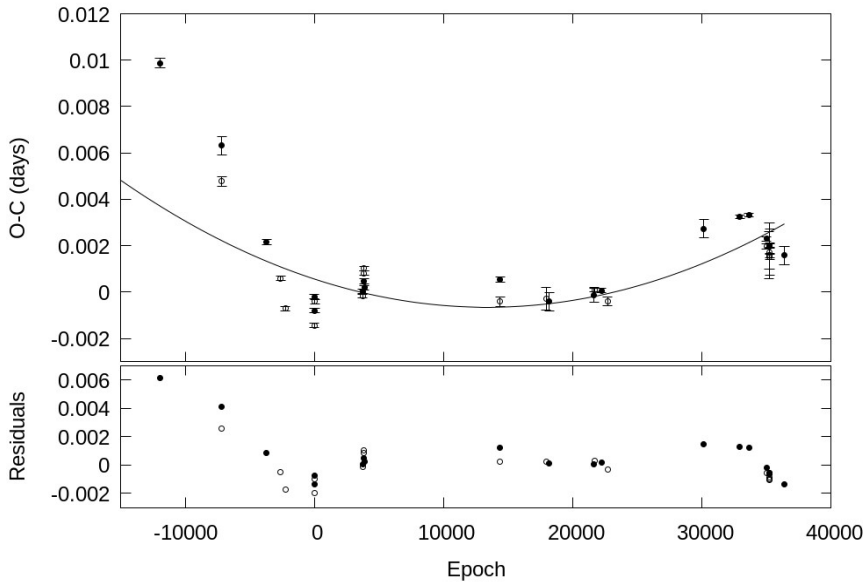

**Fig. 1** The quadratic trend on the data points and their residuals. Primary and secondary minimums are indicated as filled and open circles respectively.

As can be seen from Figure 1, the residuals are still quite large, and they do not seem to be randomly distributed. We also noted that the O-C data points could also be represented by two simple straight lines with a break at around $E \sim 0$. The first line, for points with $E < -0.5$, has a steeper downward slope, while a less pronounced ascendant slope is present for $E >= -0.5$. Proceeding with these linear fittings to each of these branches we obtained the following two ephemeris,

$$Min. I\ (BJD_{TDB}) = (2448056.89891 \pm 0.00051) + (0.30231759 \pm 1.1 \times 10^{-7}) \times E\ [days] \quad (4)$$

$$Min. I\ (BJD_{TDB}) = (2448056.90093 \pm 0.00022) + (0.302318713 \pm 9.4 \times 10^{-9}) \times E\ [days] \quad (5)$$

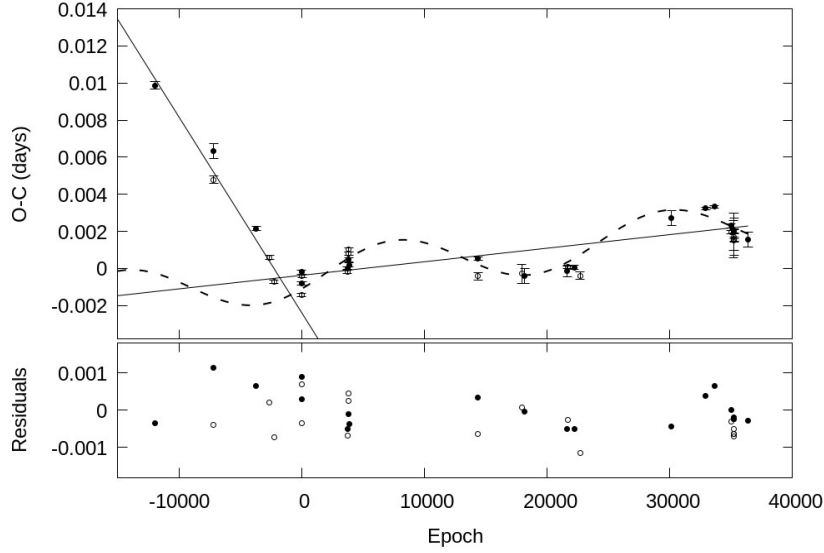

**Fig. 2** The two linear ephemeris fittings are represented as solid lines. The dashed line depicts the sinusoidal fitting to the residuals between the data and the linear ephemeris of Equation 5. In the lower panel, the final residuals after removing the linear trends and the cyclic variation are plotted. Filled and open circles represent primary and secondary minimums.

Both fittings are depicted in Figure 2. As it can be seen at the bottom of that Figure, the residuals of the first branch fitting do not show any systematic trend. The sum of the squares of the weighted residuals for each linear fitting resulted to be 1479 which is less than half of the sum of the weighted residuals of the overall quadratic fitting 3560 of Equation 2.

Each linear branch would correspond to two different constant period stages, indicating that a sudden period jump should have occurred at $E_c = -1798.75 \sim -1800$ (i.e. JD 2447513 ~ mid Dec. 1988).

The differences between the periods of Equations 4 and 5, give us the period jump which results to be $\Delta P = 1.12 \times 10^{-6} \pm 1.1 \times 10^{-7}\ days$. This period increase could be interpreted as a rapid mass transfer from the lower mass star to its companion. Supposing a conservative mass transfer the quantity of mass transferred for this period change could be derived using the following expression (e.g. Negu & Tessema 2015)

$$\frac{\Delta P}{P} = 3\Delta M \frac{(M_1 - M_2)}{M_1 M_2} \quad (6)$$

Considering $M_1$ and $M_2$ from Table 5 and the system parameters we obtain the transferred mass to be about $\Delta M = 2.45 \times 10^{-6}\ M_\odot$.

A more detailed inspection of the linear fittings of Figure 2, it looks like the data located at $E > E_c$ show an oscillatory behavior around the second branch line. Then, a new adjustment to the residuals between these data and Equation 5, was made using a sinusoidal function, to get

$$O - C = (0.0013 \pm 0.0002) \times sin((0.0002838 \pm 5.2 \times 10^{-6}) \times E + (-0.578 \pm 0.124))\ [days] \quad (7)$$

The addition of Equations 5 and 7 are depicted at the top panel of Figure 2. The residuals of the whole fittings are presented at the bottom of the same Figure. This sinusoidal variation in the O-C diagram of BF Pav presents an amplitude $K = 0.00134 \pm 0.00021$, $d = 115.5 \pm 18.1$ s and a period of $2\pi P/(0.0002838 \times 365.25) = 18.3 \pm 0.3$ yr.

Some hypotheses are commonly used to explain this kind of behavior. If a significant magnetic activity is present in one of the binary components, the changes in its inner structure during the magnetic activity cycles can cause a spin-orbital coupling producing the cyclic variation of the orbital period. This is known as the Applegate mechanism (Applegate 1992). However, as will be discussed later, BF Pav does not seem to have significant magnetic activity.

On the other hand, the periodic changes of O-C could also be attributed to the light-time effect, caused by an invisible third body revolving around the binary system (e.g. Irwin 1952). We will restrict our analyses to a circular orbit for the third body because the distribution of points of our collection of times of minimum does not deserve a more detailed model with eccentric orbits. In this case, the projected semi-axis ($a'_{12} \sin i'$) of the orbit of the binary around the barycenter of the triple system is given by

$$a'_{12} \sin i' = K \times c \quad (8)$$

where $i'$ is the inclination of the triple system orbit, K is the amplitude of the O-C oscillation (Equation 7), and c is the speed of light. Thus we obtain $a'_{12} \sin i' = 0.2324 \pm 0.037$ AU. The mass function $f(m)$ must be used to derive the projected mass of the third body ($M_3 \sin i'$),

$$f(m) = (4\pi^2/GP_3^2)(a'_{12} \sin i')^3 = (M_3 \sin i')^3/(M_1 + M_2 + M_3)^2 \quad (9)$$

where $P_3$ is the period of oscillation of Equation 6 and G is the gravitational constant.

Thereby, $f(m) = 3.69 \times 10^{-6} \pm 1.75 \times 10^{-6} M_\odot$ and the projected mass $M_3 \sin i' = 0.024 \pm 0.013 M_\odot = 25 \pm 13 M_{Jup}$. Even inside the errors, the minimum mass for the third body, for the case for $i' = 90°$, is greater or of the order of the lower limit mass for a brown dwarf ($\sim 0.014 M_\odot$).

For an inclination of $i' = 19°$, $M_3$ would correspond to a star at the lower mass limit of $0.075 M_\odot$. Supposing a uniform distribution of inclination angles, the probability of a star to have certain orbital inclination is given by the distribution function $p(i) = \sin i$. In this way, integrating $p(i)$, we obtain that the probability for the third body to be a low mass star with $M >= 0.075 M_\odot$ is 5.4%. However, the third body has many more chances to be a brown dwarf with a probability of 94.6%. Using the third Kepler law $a_3[AU] = ([P_3^2[yr] \times (M_1 + M_2 + M_3)[M_\odot]])^{1/3}$ we can derive the semi-axis of the third body orbit. Supposing an orbital inclination $i' = 90°$, $a_3 = 8.04 \pm 0.33$ AU, which is quite larger than the common envelope of the binary.

**4. LIGHT CURVE ANALYSIS**

We used the Wilson-Devinney code (Wilson & Devinney 1971), to analyze the light curves. We preferred to use the W-D code combined with the Monte Carlo simulation to determine the uncertainties of the adjustable parameters (Zola et al. 2004, 2010). The mass ratio of the system could be obtained by the *q*-search method in the photometric observations, so we did it according to the required standards (Rucinski et al. 2005).

The (*B-V*) color index is the difference in magnitudes between two wavelength filters *B* and *V*. The blue and visual magnitudes are measured through filters centered at 442 nm and 540 nm, respectively. Passing light through different filters depends on the star's surface temperature according to the Planck Law radiation distributions. It means that by having data of the Blue and Visual filters we can calculate the (*B-V*) index and obtain a good estimation of a star's surface temperature by it (Poro et al. 2020).

The fraction of detected flux of wavelength depends on the telescope mirrors, the bandwidth of filters, and the response of the photometer, thus it is necessary to correct our data by calibration with the comparison stars from standard catalogs.

Many studies presented relations between the (*B-V*) index and the surface temperature of the star such as Code et al. (1976), Sekiguchi and Fukugita (2000), and Ballesteros (2012). Eker et al. (2018) presented relations and tables for different parameters of the main-sequence stars. Eker et al. (2018) selected absolute parameters of 509 main-sequence stars from the components of detached-eclipsing spectroscopic binaries in the solar neighborhood that are used to study Mass-Luminosity (*ML*), Mass-Radius (*MR*), and Mass-Temperature (*MT*) relations. They combined the photometric data of Sejong Open cluster Survey (SOS) and typical absolute parameters adjusted from the *ML*, *MR*, and *MT relation* functions calibrated in their study. 'Sejong Open cluster Survey (SOS) is a photometry project of a large number of clusters in the SAAO Johnson-Cousins' *UBVI* system by Sung et al. (2013).

Based on our data and after calibrating (Høg et al. 2000), we calculated (*B-V*)$_{BF\ Pav}$ = $0^m.803$. As a result, based on Eker et al. (2018), the effective temperature of the secondary component was found to be 5201 $K$.

Sekiguchi and Fukugita (2000) derived a (*B-V*) color-temperature relation too. They present $T_{eff}$ as a function of (*B-V*) color index to represent the metallicity value in four classes. By combining the previous results from Eker et al. (2018) and exerting the results of Sekiguchi and Fukugita (2000), the metallicity (Fe/H) value for the primary component of BF Pav can be estimated between -0.75 and -0.25 (star's population II). As shown in Figure 3, the obtained temperature from derived (*B-V*) color is also in an acceptable range (4800 $K$ - 5300 $K$) for the primary component of BF Pav with the method of Sekiguchi and Fukugita (2000).

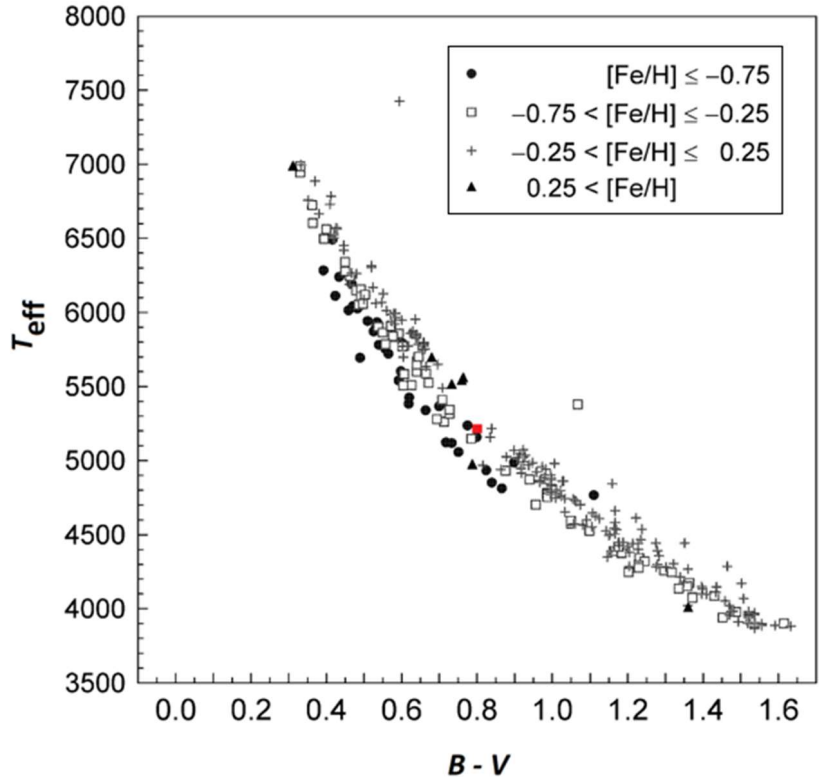

**Fig. 3** BF Pav's position (red dot) based on the Sekiguchi and Fukugita (2000) results.

We assumed gravity-darkening coefficients $g_1 = g_2 = 0.32$ (Lucy 1967), bolometric albedo $A_1 = A_2 = 0.5$ (Rucinski 1969), and linear limb darkening coefficients taken from tables published by Van Hamme (1993) in the light curve analysis.

As can be inferred from the light curves, the mean minimum occurred first, and also the temperature of the primary star is higher than the secondary. Based on the unequal minimums, and the logical light curve solutions, mode 3 was chosen for analysis. The parameters obtained from the solutions are given in Table 3. The mean fractional radii of components were calculated with the formula, *r*$_{mean}$ = (*r*$_{back}$ × *r*$_{side}$ × *r*$_{pole}$)$^{1/3}$ (10).

Table 3 Photometric solutions of BF Pav.

| Parameter | This study | Gonzales et. al. (1996) |
|---|---|---|
| $T_1$ (K) | 5420(6) | 5430 |
| $T_2$ (K) | 5201 | 5330(20) |
| $\Omega_1=\Omega_2$ | 4.394(21) | 4.320 |
| $i$ (deg) | 87.97(45) | 84.8(1.0) |
| $q$ | 1.460(14) | 1.4(2) |
| $l_1/l_{tot}(B)$ | 0.471(4) | 0.450(30) |
| $l_2/l_{tot}(B)$ | 0.529(5) | 0.550 |
| $l_1/l_{tot}(V)$ | 0.464(4) | 0.445(30) |
| $l_2/l_{tot}(V)$ | 0.536(5) | 0.555 |
| $l_1/l_{tot}(R)$ | 0.457(4) | |
| $l_2/l_{tot}(R)$ | 0.543(5) | |
| $l_1/l_{tot}(I)$ | 0.450(4) | |
| $l_2/l_{tot}(I)$ | 0.550(5) | |
| $A_1 = A_2$ | 0.50 | 0.5 |
| $g_1 = g_2$ | 0.32 | 0.32 |
| $f$ (%) | 12.5 (3.0) | 10 |
| $r_1$(back) | 0.385(2) | 0.386(15) |
| $r_1$(side) | 0.349(2) | 0.350(15) |
| $r_1$(pole) | 0.332(2) | 0.334(15) |
| $r_2$(back) | 0.452(3) | 0.445(15) |
| $r_2$(side) | 0.419(3) | 0.412(15) |
| $r_2$(pole) | 0.395(3) | 0.390(15) |
| $r_1$(mean) | 0.355(2) | 0.355(15) |
| $r_2$(mean) | 0.421(3) | 0.414(15) |
| Colatitude$_{spot}$ (deg) | 25(4) | |
| Longitude$_{spot}$ (deg) | 120(2) | |
| Radius$_{spot}$ (deg) | 39(2) | |
| $T_{spot}/T_{star}$ | 0.90(2) | |
| Phase Shift | -0.0179(1) | |

Note: Parameters of a star spot is on the secondary component.

The observed and synthetic light curves in *BVRI* filters with residuals show in Figure 4.

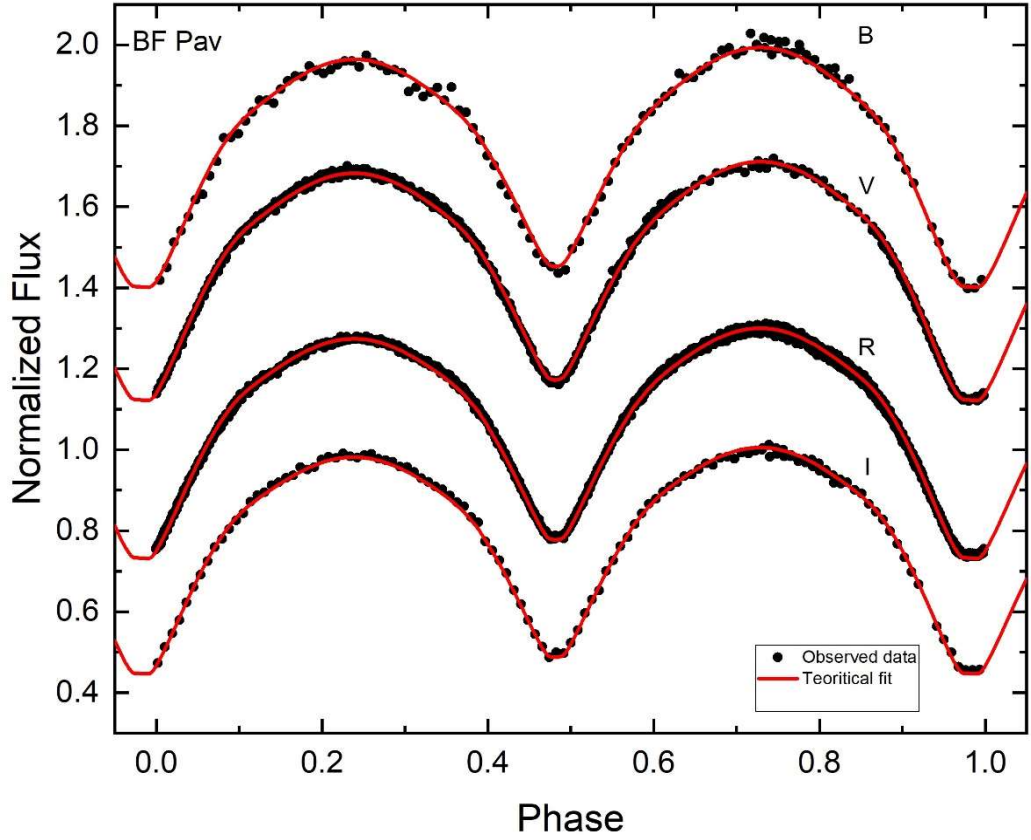

**Fig. 4** Observed light curves of BF Pav (points) and modeled solutions (lines) in the *BVRI* filter from top to bottom, respectively, and residuals are plotted; with respect to orbital phase, shifted arbitrarily in the relative flux.

Fillout factor is a quantity that indicates the degree of contact in the binary star systems defined by Mochnacki and Doughty (1972), and also Lucy and Wilson (1979) that was modified and redefined by Bradstreet (2005),

$$f = \frac{\Omega(L_1) - \Omega}{\Omega(L_1) - \Omega(L_2)} \qquad (11)$$

where $\Omega$, $\Omega(L_1)$, and $\Omega(L_2)$ are star surface potential, inner Lagrangian surface potential, and outer Lagrangian surface potential, respectively. We calculated a fillout factor of 12.5% with a cold spot from the output parameters of the light curve solutions.

A difference in the heights of the maxima in light curves of eclipsing binary systems indicates the O'Connell effect (O'Connell 1951). This binary system appears to demonstrate this effect because we need to add a spot on the secondary component in the light curve solutions. Table 4 represents the characteristic parameters of the light curves of BF Pav.

**Table 4** Characteristic parameters of the light curves in the *BVRI* filters.

| Part of LC. | B | V | R | I |
|---|---|---|---|---|
| MaxI - MaxII | 0.032 | 0.005 | 0.027 | 0.011 |
| MaxI - MinI | -1.052 | -0.982 | -0.920 | -0.879 |
| MaxI - MinII | -0.868 | -0.820 | -0.783 | -0.780 |
| MinI - MinII | 0.184 | 0.162 | 0.137 | 0.099 |

The $M_{secondary}$ is derived from a study by Eker et al. (2018), and $M_{primary}$ is calculated by $q = \frac{M_2}{M_1}$. We also calculated the mass of each component of the binary system using the method of Harmanec (1988) who derived a simple approximation formula relating absolute parameters (mass, radius, and luminosity) to the effective temperature of the components based on data analysis. For this purpose, we used the following formula,

$$log \frac{M}{M_\odot} = ((1.771141X - 21.46965)X + 88.05700)X - 121.6782 \quad (12)$$

where $X$ is $log(T_{eff})$. This formula is only defined in the range of $4.62 \geq log(T_{eff}) \geq 3.71$ (Harmanec 1988). So we calculated M$_2$ as the mentioned range is valid for secondary $T_{eff}$ of BF Pav due to our photometric solution. The absolute parameters are given in Table 5 and there is high conformity between the results which were obtained by two methods.

Table 5  Estimated absolute parameters of BF Pav by two methods to calculate the mass of the primary component.

| Parameter | Eker et al. (2018) | | Harmanec (1988) | |
|---|---|---|---|---|
| | Primary | Secondary | Primary | Secondary |
| Mass ($M_\odot$) | 0.626(21) | 0.914(32) | 0.621(18) | 0.906(18) |
| Radius ($R_\odot$) | 0.795(8) | 0.903(10) | 0.792(1) | 0.900(1) |
| Luminosity ($L_\odot$) | 0.488 (57) | 0.534(65) | 0.485(57) | 0.531(65) |
| $M_{bol}$ (mag) | 5.53(14) | 5.43(14) | 5.53(14) | 5.44(14) |
| $log\ g$ (cgs) | 4.43(2) | 4.49(2) | 4.43(2) | 4.49(2) |
| $a(R_\odot)$ | 2.24(1) | | 2.23(1) | |

According to the estimated absolute parameters of this binary system, the distance was calculated. We obtained $m_{system} = 12.908(25)$ from our light curve and $M_v = 5.611(29)$ for the secondary component ($BC_1 = -0.181$ from Eker et al. 2018). So the distance to the binary system computed from the formula,

$$d_{(pc)} = 10^{\left(\frac{m_{system}-M_{pri}+5-A_v}{5}\right)} \quad (13)$$

Therefore, an estimate of the distance of this binary system is $268 \pm 18$ parsec (using $A_v = 0.155$ (Schlafly & Finkbeiner 2011).

The 3D view of BF Pav and the Roche lobe configuration of BF Pav illustrated in Figure 5.

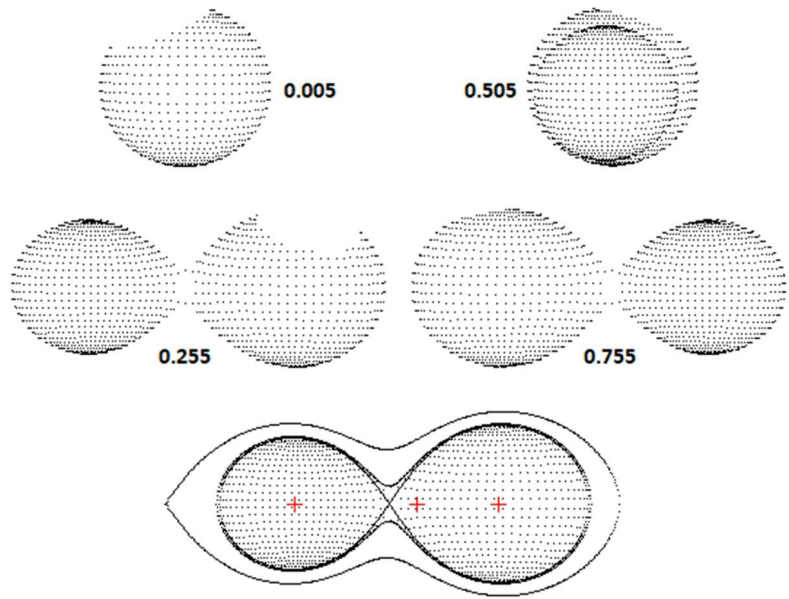

Fig. 5 The positions of the components of BF Pav.

## 5. RESULT AND CONCLUSION

The photometric observations of BF Pav were carried out during eight nights utilizing *BVRI* filters. This study's approach is to present a new ephemeris and light curve analysis of the W UMa-type eclipsing binary BF Pav and probe this binary system's period changes.

According to a quadratic trend in the O-C diagram, we obtain a period increase at a rate of $(16.4 \pm 3.6) \times 10^{-8}$ d/yr or 0.142 s/century, but a probably sudden period jump $\Delta P = 1.1 \times 10^{-7}$ days could have occurred instead at $E_c$ 1800. This period increase could be interpreted as a rapid mass transfer from the lower mass star to its companion. Supposing a conservative mass transfer the quantity of mass transferred for this period change can be derived to be about $\Delta M = 2.45 \times 10^{-6} M_\odot$. The data located at $E > E_c$ show an oscillatory behavior around the second branch line. This variation has a period of $18.3 \pm 0.3$ yr in the O-C diagram. Since this system does not seem to have significant magnetic activities, this cyclic trend could be attributed to the light-time effect caused by an invisible third body in the system. We obtain that the probability for the third body to be a low mass star with $M \geq 0.075 M_\odot$ is 5.4%. While, the third body has many more chances to be a brown dwarf with a probability of 94.6%. Supposing $i' = 90°$ the semi-major axis of the third body becomes $a_3 = 8.04 \pm 0.33$ AU, which is quite larger than the common envelope of the binary. This system should be followed up by other future observations and more times of minima to reveal the nature of orbital period variations and our detected cycle in it. Hence these models can be considered as speculations for future reference.

We specified the photometric solution of the short period system BF Pav based on the Wilson-Devinney code combined with the MC simulation to calculate the uncertainties of the searched parameters. We obtained a mass ratio ($q = \frac{M_2}{M_1}$) of $1.460 \pm 0.014$ from the *q*-search method which suggested that BF Pav is a contact binary, a fillout factor (*f*), and an inclination (Table 3). Also, the difference between this binary system components' temperature $\Delta T$ in the order of 200 $K$. We calculated the binary system distance which equals to $268 \pm 18\ pc$ and this result is in good agreement with the Gaia EDR3 value $253.272 \pm 0.992\ pc$.

Based on the estimation of absolute parameters, the diagrams of the Mass-Luminosity (*M-L*) and the Mass-Radius (*M-R*) on a *log*-scale show the evolutionary status of BF Pav. The theoretical ZAMS and TAMS lines and the positions of the primary and secondary components are depicted in Figure 6. Since the W UMa-type eclipsing binaries are known as the Low-Temperature Contact Binaries (LTCBs), the difference between the temperatures of two components are close to each other and typically around 5%; and this is about 4% for BF Pav. As discussed by Yakut and Eggleton (2005), in this type of contact binary system the luminosity of some primary is transferred to the secondary because of their initial masses. Moreover, in W UMa-type eclipsing binaries the components share a common convective envelope, so the primary component is near to the zero-age main sequence (Figure 8-a). This is taken to mean that the primary is not yet evolved. Alternatively, the deviation of the secondary component shows it is slightly evolved from ZAMS. The secondary components show different evolutionary paths due to more initial masses than the present masses (Yildiz & Doğan 2013). According to the amount of the mass ratio, the fillout factor, which complies with Gonzalez et al. (1996) results, we suggest that BF Pav is a W-type system. Yildiz and Doğan (2013) investigated the parameters of W-type W UMa binaries to estimate initial masses of these stars which were obtained based on MESA models (Paxton et al. 2010) due to mass transfer between two components. According to the mass loss model of Yildiz and Doğan (2013) and clearly from Figure 6, on the *log M - log L* diagram, the location of both components of BF Pav appearance to be in good agreement with the distribution of primary and secondary stars of the W-type W UMa binary systems.

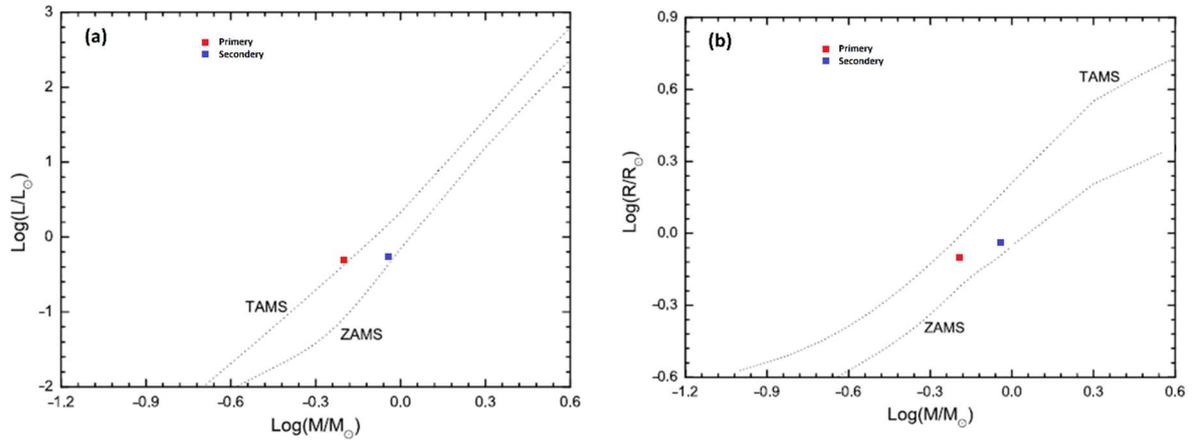

**Fig. 6** The *log M - log L*, and *log M - log R* diagrams for BF Pav from the absolute parameters. The dashed lines represent the TAMS and ZAMS and the locations of the primary and secondary components of BF Pav are marked.

Stellar winds are the major responsible for the binary system's mass loss due to the star's magnetic activities. According to Table 4 and maxima differences in the light curves, it seems that BF Pav does not have significant magnetic activity and this implies a negligible O'Connell effect in this binary system and we fetched up that the mass loss idea is void for this system, so we concentrate on mass transfer.

BF Pav had been observed by Hoffman (1981) but the observer has not been able to prosecute a detailed analysis due to lack of data. After a while, the first detailed photometric analysis of BF Pav was performed in 1996 using the Wilson-Devinney code (Gonzalez et al. 1996) after *UBV* photoelectric observations of this binary system between 1987 and 1993 in the observational program of southern short-period eclipsing binaries. The former photometric solution demonstrates that BF Pav has a mass ratio of 1.4 while we calculated $q = 1.460 \pm 0.014$. To complete our comparison, we found 12.5% for the amount of fillout factor whereas 10% was obtained for *f* in the prior study.

**Acknowledgments** This manuscript is based on the Binary Systems of South and North Project (http://bsnp.info). The project aims to study contact binary systems in the northern and southern hemispheres of several observatories in different countries.